# Time-stretch mid-infrared spectroscopy with undetected photons

Makoto Shoshin[1,†], Kazuki Hashimoto[2,3,4,†,*], Tatsuki Yamazumi[1], and Takuro Ideguchi[1,2,3,4,**]

[1]Department of Physics, The University of Tokyo, Tokyo 113-0033, Japan

[2]Institute for Photon Science and Technology, The University of Tokyo, Tokyo 113-0033, Japan

[3]Department of Advanced Materials Science, The University of Tokyo, Chiba, 277-0882, Japan

[4]RIKEN Center for Advanced Photonics, RIKEN, Saitama, 351-0198, Japan

[†]These authors contributed equally to this work.

*k-hashimoto@ipst.s.u-tokyo.ac.jp

**ideguchi@g.ecc.u-tokyo.ac.jp

**Abstract**

Mid-infrared (MIR) spectroscopy probes fundamental molecular vibrations, providing insights into chemical composition and changes in molecular environments. However, high-speed measurements remain challenging because of the limited sensitivity and bandwidth of MIR photodetectors. Spectroscopy with undetected photons (SUP) offers an alternative by exploiting parametric down-conversion, which generates frequency-correlated MIR and near-infrared (NIR) light. Through nonlinear interference, MIR absorption information is encoded in the NIR spectrum, eliminating the need for MIR photodetection. Nevertheless, MIR-SUP has been limited by low photon flux or slow spectral readout. Here, we overcome these limitations by combining high-gain SU(1,1) interferometry with time-stretch spectroscopy. Phase-sensitive amplification produces bright NIR signal pulses, enabling pulse-resolved spectral detection using telecom-band components. Driven by a fiber-amplified gain-switched laser diode, our system achieves millisecond-scale MIR transmittance spectroscopy of liquid benzene near 3.2 μm, accelerating spectral acquisition compared with previous MIR-SUP demonstrations. We further demonstrate broadband measurements spanning approximately 130 cm$^{-1}$ using a shorter nonlinear crystal. By combining nonlinear interferometric sensing with ultrafast spectral readout, our approach establishes a route toward high-speed MIR spectroscopy using a simple semiconductor-laser-based platform.

**Introduction**

Mid-infrared (MIR) spectroscopy provides chemically specific information by probing fundamental molecular vibrations, enabling analysis of molecular composition and changes in local environments. High-speed MIR spectroscopy is particularly valuable for monitoring rapid and nonrepetitive processes, such as combustion and fluid flows [1–4], and for increasing measurement throughput in applications such as hyperspectral imaging [5,6]. However, conventional broadband MIR spectrometers, such as Fourier-transform infrared (FTIR) spectrometers [7], typically rely on low-brightness thermal light sources and direct MIR photodetection. The comparatively low sensitivity and limited bandwidth of MIR photodetectors pose challenges for high-speed measurements.

Spectroscopy with undetected photons (SUP) offers an alternative that eliminates the need for direct MIR photodetection by exploiting nonlinear interference [8,9]. This technique uses parametric down-conversion (PDC) to generate broadband, frequency-correlated visible/near-infrared (VIS/NIR) signal and MIR idler light from a narrowband pump source. By allowing the MIR idler light to interact with a sample between two nonlinear interactions, the sample's absorption information is encoded in the interference fringes of the signal spectrum, enabling MIR spectroscopy through VIS/NIR detection. However, most previous MIR-SUP implementations have relied on spontaneous parametric down-conversion (SPDC) in the low-gain regime, where the mean photon number per mode is much less than unity [10–15]. The resulting low photon flux has typically limited spectral acquisition rates to the mHz–Hz range [10–15], although a recent demonstration achieved 100 Hz [15]. Alternatively, high-gain PDC has been employed to generate bright signal–idler twin beams, overcoming this photon-flux limitation [16]. Nevertheless, the demonstrated spectral acquisition rates with scanning Fourier-transform spectral readout have remained in the mHz range.

Here, we develop time-stretch MIR-SUP (TS-MIR-SUP) by combining high-gain SU(1,1) interferometry [17], a form of nonlinear interferometry, with time-stretch spectroscopy [18]. In the second nonlinear interaction of the interferometer, phase-sensitive amplification (PSA) produces spectral interference fringes in the NIR signal, whose visibility depends on MIR absorption in the idler path. The amplification also increases the NIR signal intensity sufficiently for pulse-resolved spectral detection. Time-stretch spectroscopy maps each signal spectrum onto a temporal waveform through chromatic dispersion, enabling high-speed readout using telecom-band optical fibers and photodetectors. The MIR transmittance spectrum is retrieved from the fringe visibility extracted after averaging multiple pulse-resolved spectra. Importantly, high-gain operation is achieved using a fiber-amplified gain-switched laser diode as the sole pump source, without requiring a broadband MIR laser. Using this system, we demonstrate millisecond-scale MIR transmittance spectroscopy of liquid benzene near 3.2 μm, achieving a signal-to-noise ratio (SNR) of approximately 24 with 1,000-pulse averaging at a repetition rate of 1 MHz. We further demonstrate broadband transmittance measurements of a bandpass filter and an acrylic plate over a spectral range of approximately 130 $cm^{-1}$ near 2.9 μm using a shorter nonlinear crystal.

## Results

### Principle of TS-MIR-SUP

Figure 1 illustrates the principle of TS-MIR-SUP, which combines high-gain SU(1,1) interferometry with time-stretch spectroscopy. The SU(1,1) interferometer consists of two successive nonlinear interactions, implemented either in separate nonlinear media (Figure 1) or through two passes through the same medium in a folded configuration (Figure 2). In the first interaction, a picosecond pump pulse drives PDC, generating broadband, frequency-correlated NIR signal and MIR idler beams. Their spectral correlations are governed by energy conservation under narrowband pumping. An optical delay $\tau$ between the pump–signal and idler paths determines the spectral fringe frequency, while the MIR idler beam passes through the sample and undergoes wavelength-dependent attenuation and phase shifts.

In the second nonlinear interaction, the pump drives PSA of the correlated signal and idler fields. Depending on their relative phases, the signal field is amplified or deamplified, producing spectral interference fringes in the output NIR signal. MIR absorption in the idler path modifies the fringe visibility, thereby encoding the sample's transmittance spectrum in the NIR signal without requiring MIR photodetection. This interference is the high-gain counterpart of that observed in low-gain SU(1,1) interferometry, where indistinguishable photon-pair generation processes give rise to interference in the detected photon counts [19]. Importantly, PSA also increases the signal pulse energy by four to five orders of magnitude relative to that after the first PDC process, reaching the pJ level and enabling pulse-resolved detection with conventional photodetectors.

The NIR signal spectrum containing the interference fringes is read out using time-stretch spectroscopy. Propagation through a dispersive medium, such as an optical fiber, maps the spectrum of each signal pulse onto a temporally stretched waveform, which is recorded by a high-speed photodetector. Although individual pulse spectra can be detected, multiple spectra are averaged to suppress intensity fluctuations and reliably extract the fringe visibility. The MIR transmittance spectrum is then retrieved from the fringe visibilities measured with and without the sample, accounting for their nonlinear dependence on idler attenuation [20].

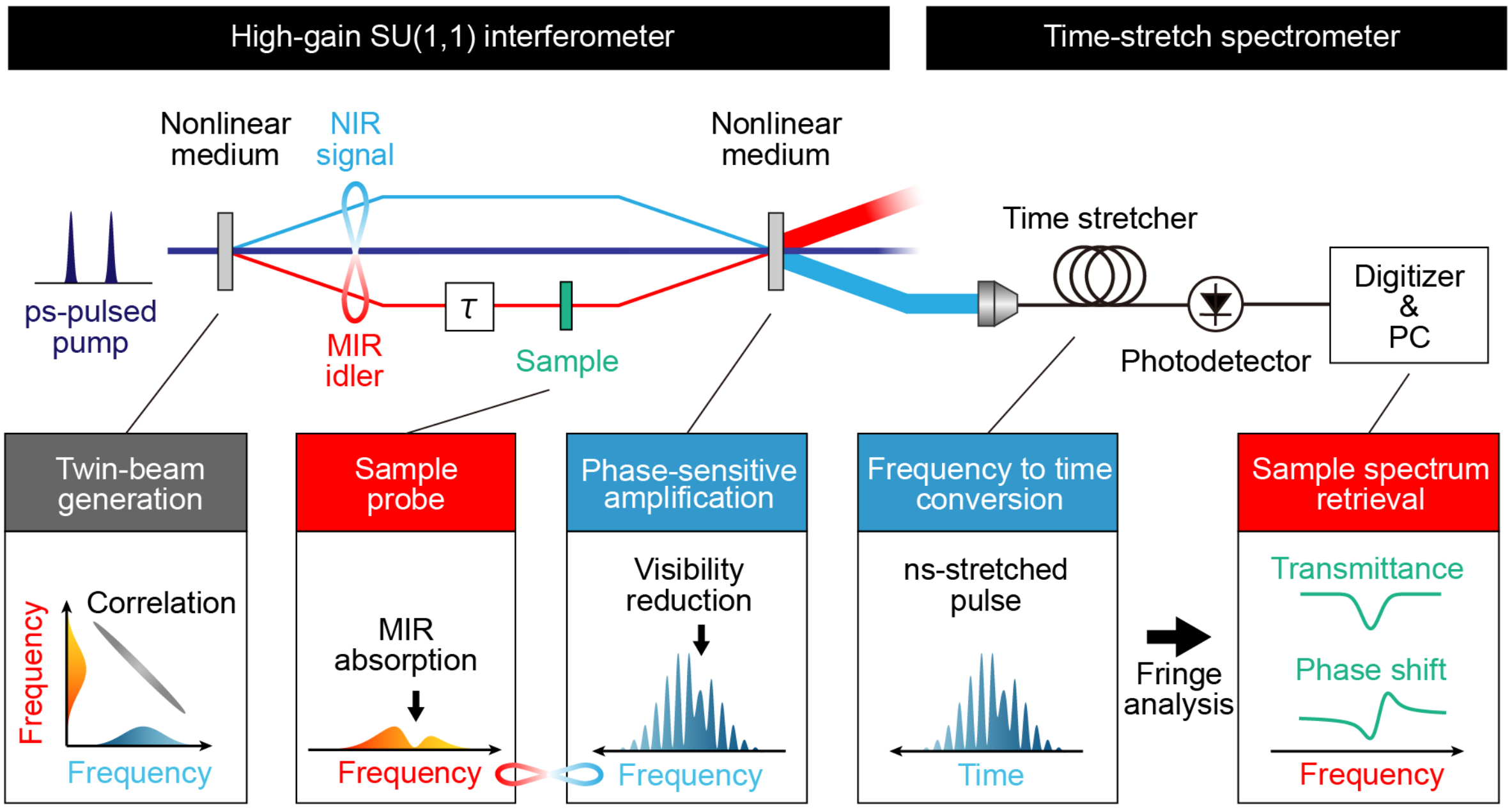

**Fig. 1: Principle of TS-MIR-SUP.** Frequency-correlated NIR signal and MIR idler beams are generated by PDC. After the idler beam interacts with the sample, PSA in the second nonlinear interaction produces spectral interference fringes in the NIR signal, whose visibility depends on the MIR transmittance. The NIR signal spectrum is subsequently read out by time-stretch detection. The optical delay $\tau$ determines the spectral fringe frequency.

## Experimental setup

Figure 2 shows the experimental implementation of TS-MIR-SUP. The setup is outlined below, with further details provided in Methods. The pump source was a gain-switched laser diode generating 50-ps pulses centered at 1029 nm. Direct electrical modulation of the semiconductor laser enables picosecond-pulse generation with a compact, robust, and cost-effective configuration and a widely tunable repetition rate [21]. The pulses were subsequently amplified to sub-μJ energies using ytterbium-doped fiber amplifiers (YDFAs) to drive high-gain PDC. The amplified pump pulses were focused into a periodically poled lithium niobate (PPLN) crystal, generating NIR signal and MIR idler beams centered around 1.5 and 3 μm, respectively. Two PPLN crystals with different lengths were used to balance parametric gain and spectral bandwidth. For high-speed measurements, a 50-mm-long crystal provided sufficient gain at relatively low pump pulse energies, allowing operation at a repetition rate of 1 MHz. For broadband measurements, a shorter, 10-mm-long crystal provided a wider phase-matching bandwidth, while the repetition rate was reduced to 50 kHz to maintain sufficient pulse energy for high-gain operation. The parametric gain parameters for both configurations were evaluated as described in Supplementary Note 1.

The SU(1,1) interferometer was implemented in a folded configuration, in which the pump and signal beams were reflected directly back into the same crystal, while the idler beam was reflected after passing through the sample. The optical-path length difference between the pump–signal and idler arms determines the spectral fringe spacing and, after time-stretch mapping, the RF fringe frequency. This frequency was selected to allow separation of the interference fringes from the slowly varying spectral background while avoiding excessive attenuation by the finite temporal correlation response.

After the second passage through the crystal, the NIR signal pulses, with energies on the order of pJ, were directed to the time-stretch detection system comprising telecom-band optical fibers and photodetectors. The chromatic dispersion was selected such that the temporal correlation width of the stretched signal was sufficiently larger than the detector impulse-response width, minimizing the influence of the detector response on the spectral resolution. Under these conditions, further increasing the dispersion would introduce additional fiber loss without substantially improving the effective spectral resolution.

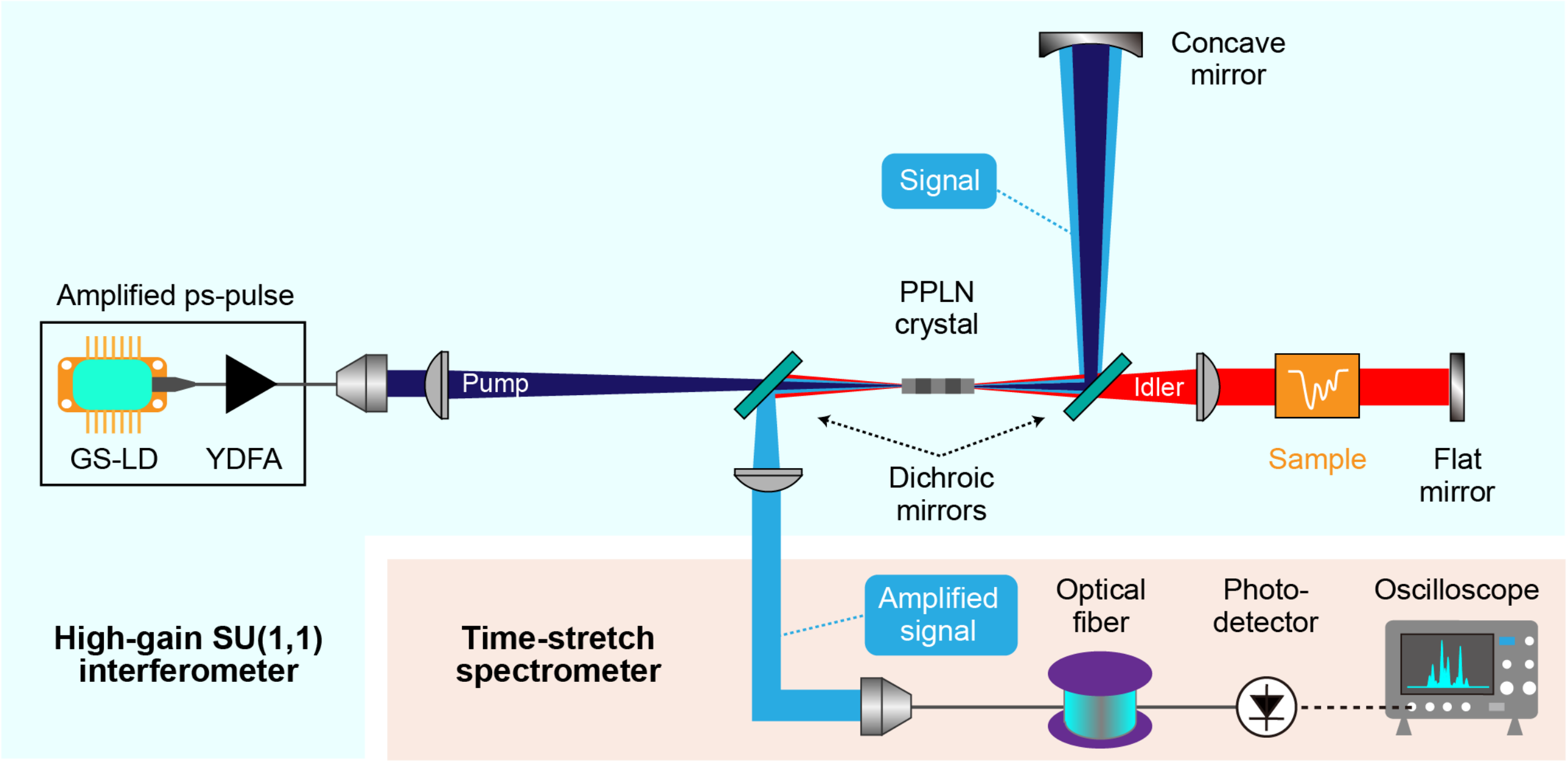

**Fig. 2: Experimental implementation of TS-MIR-SUP.** A gain-switched laser diode followed by a YDFA provides picosecond pump pulses for high-gain PDC in a PPLN crystal. The SU(1,1) interferometer is implemented in a folded configuration, with the MIR idler beam passing through the sample. The amplified NIR signal is subsequently directed to the time-stretch spectrometer for pulse-resolved spectral detection. GS-LD, gain-switched laser diode; YDFA, ytterbium-doped fiber amplifier; PPLN, periodically poled lithium niobate.

**Pulse-to-pulse characterization of time-stretch waveforms**

We first characterized the pulse-to-pulse intensity fluctuations of the time-stretch signal. Figure 3a shows a heatmap of 4,000 consecutive waveforms acquired at a repetition rate of 1 MHz using the 50-mm-long PPLN crystal, corresponding to a total measurement time of 4 ms. The horizontal axis represents the intra-pulse time, which is mapped to optical wavelength through time-stretch detection, while the vertical axis represents the inter-pulse time. A clear interference fringe pattern is visible across the consecutive waveforms, despite substantial intensity fluctuations in individual pulses. Figure 3b shows representative single-shot waveforms, revealing pronounced intensity fluctuations both between pulses and across spectral components within each pulse.

To quantify these fluctuations, we first examined the intensity statistics at a fixed spectral component using 8,000 single-shot waveforms. Figure 3c shows a histogram of the signal intensity at an intra-pulse time of 0 ns, with the bin width determined using the Freedman–Diaconis rule [22]. The positive-intensity region is well described by an exponential distribution, consistent with the thermal photon-number statistics of high-gain PDC light [23]. The distribution extends slightly into negative intensities due to the contribution of Gaussian detection noise. Figure 3d shows the SNR evaluated at a fixed intra-pulse time as a function of the number of averaged waveforms, $N$. For each $N$, the 8,000 intensity values were divided into groups of $N$ consecutive waveforms, and the SNR was calculated as the ratio of the mean to the standard deviation of the group-averaged intensities. The measured SNR is approximately equal to $\sqrt{N}$, as expected from thermal statistics.

We next examined correlations between intensity fluctuations at different spectral components. Figure 3e shows a two-dimensional correlation map calculated from the same 8,000 single-shot waveforms using the Pearson correlation coefficient. The pronounced diagonal ridge indicates correlated intensity fluctuations between neighboring spectral components. A cross-section at an intra-pulse time of 0 ns exhibits a full width at half maximum (FWHM) of 0.67 ns, corresponding to a wavenumber width of 1.7 $cm^{-1}$ based on the measured chirp rate of 2.5 $cm^{-1}$ $ns^{-1}$ at 1.51 μm (see Methods). This width is comparable to the pump spectral linewidth of 1.3 $cm^{-1}$ with some broadening expected in the high-gain regime [24]. The periodic modulation along the diagonal ridge originates from the spectral interference fringes of the SU(1,1) interferometer, as further discussed in Supplementary Note 2.

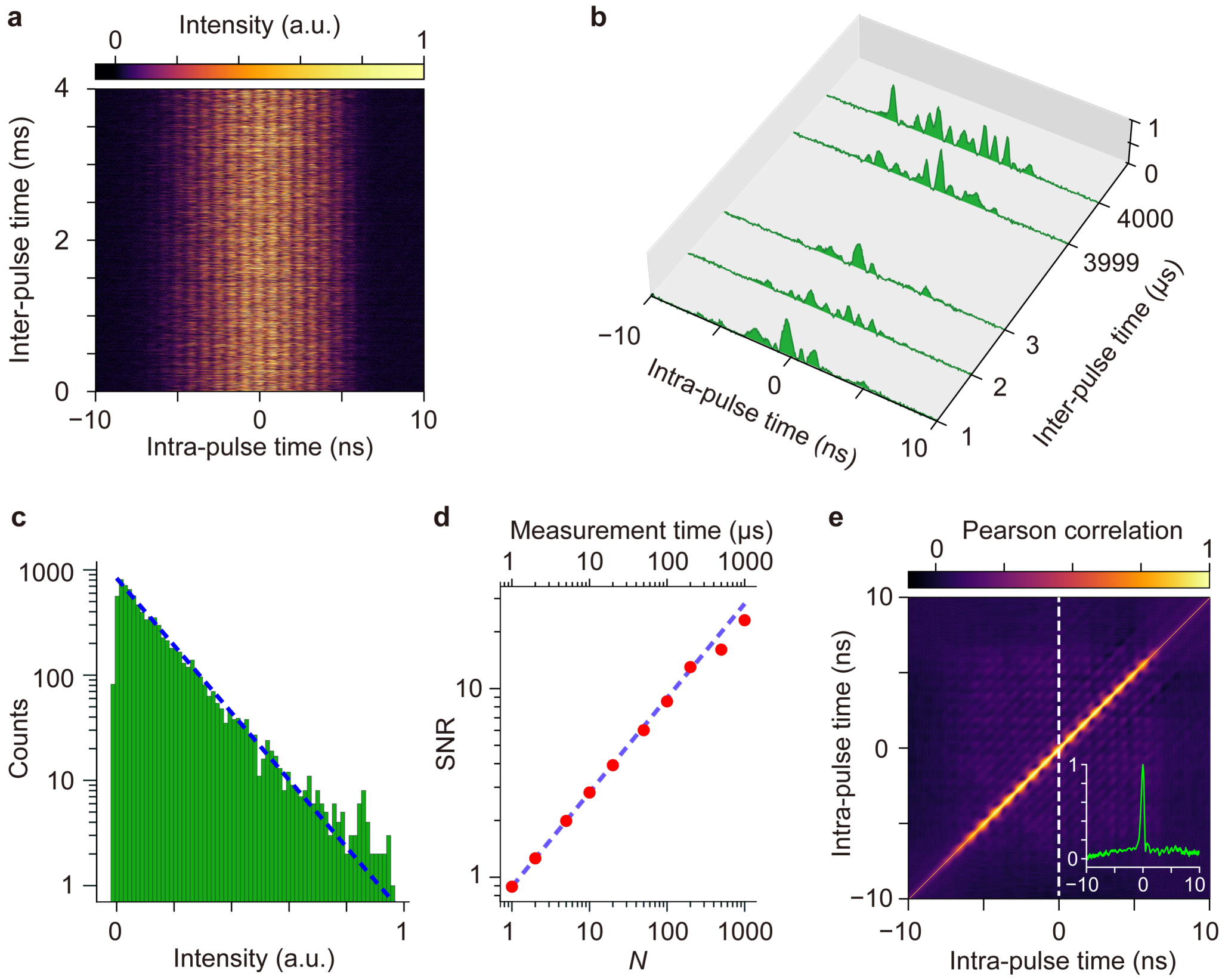

**Fig. 3: Pulse-to-pulse characterization of time-stretch waveforms.** (a) Heatmap of 4,000 consecutive time-stretch waveforms acquired at a repetition rate of 1 MHz. (b) Representative consecutive single-shot waveforms showing pulse-to-pulse intensity fluctuations. (c) Histogram of the signal intensity at an intra-pulse time of 0 ns obtained from 8,000 waveforms. The dashed line represents an exponential fit. (d) SNR as a function of the number of averaged waveforms, $N$. The dashed line represents a fit to the experimental data for $N \leq 200$, given by $0.94\sqrt{N}$. (e) Two-dimensional correlation map of signal intensities at different intra-pulse times, calculated using the Pearson correlation coefficient. The inset shows a cross-section at 0 ns, indicated by the white dotted line, with a FWHM of 0.67 ns.

**High-speed MIR transmittance spectroscopy**

To demonstrate high-speed MIR spectroscopy, we measured the transmittance spectrum of liquid benzene using the 50-mm-long PPLN crystal at a pump repetition rate of 1 MHz. Figure 4a shows the time-stretch signal spectra measured with and without the sample, each averaged over 1,000 consecutive pulses. The intra-pulse time axis was calibrated to the signal wavenumber axis. Figure 4b shows the corresponding normalized interference waveforms, obtained by dividing each measured waveform by its non-interference background extracted using a digital low-pass filter. The fringe visibilities were determined from the envelopes of the normalized waveforms using the Hilbert transform. The absorption features of benzene appear as reductions in fringe visibility.

Figure 4c shows the MIR transmittance spectrum retrieved from the fringe visibilities measured with and without the sample. A total of 8,000 waveforms were divided into eight groups of 1,000 pulses, yielding eight independently retrieved transmittance spectra from which the mean and standard deviation were calculated. Owing to the frequency correlation between the signal and idler beams, the MIR idler wavenumber was obtained by subtracting the signal wavenumber from the pump wavenumber. The retrieved spectrum over a 20 cm$^{-1}$ range agrees well with the reference FTIR spectrum convolved to the estimated spectral resolution of 6.2 cm$^{-1}$. With 1,000-pulse averaging at a repetition rate of 1 MHz, each transmittance spectrum is obtained with an acquisition time of 1 ms, corresponding to a spectral acquisition rate of 1 kSpectra/s.

Figure 4d shows the SNR as a function of the number of averaged pulses, $N$. The SNR was calculated as the inverse of the root-mean-square of the standard deviations of independently retrieved transmittance spectra across the measured spectral range. The measured SNR increases approximately as $\sqrt{N}$, reaching 24 with 1,000-pulse averaging. In addition to transmittance, the spectral phase shift induced by liquid benzene can be retrieved from the same time-stretch measurements, as demonstrated in Supplementary Note 3. Further details of the spectral calibration, fringe-visibility extraction, and transmittance retrieval are provided in Methods.

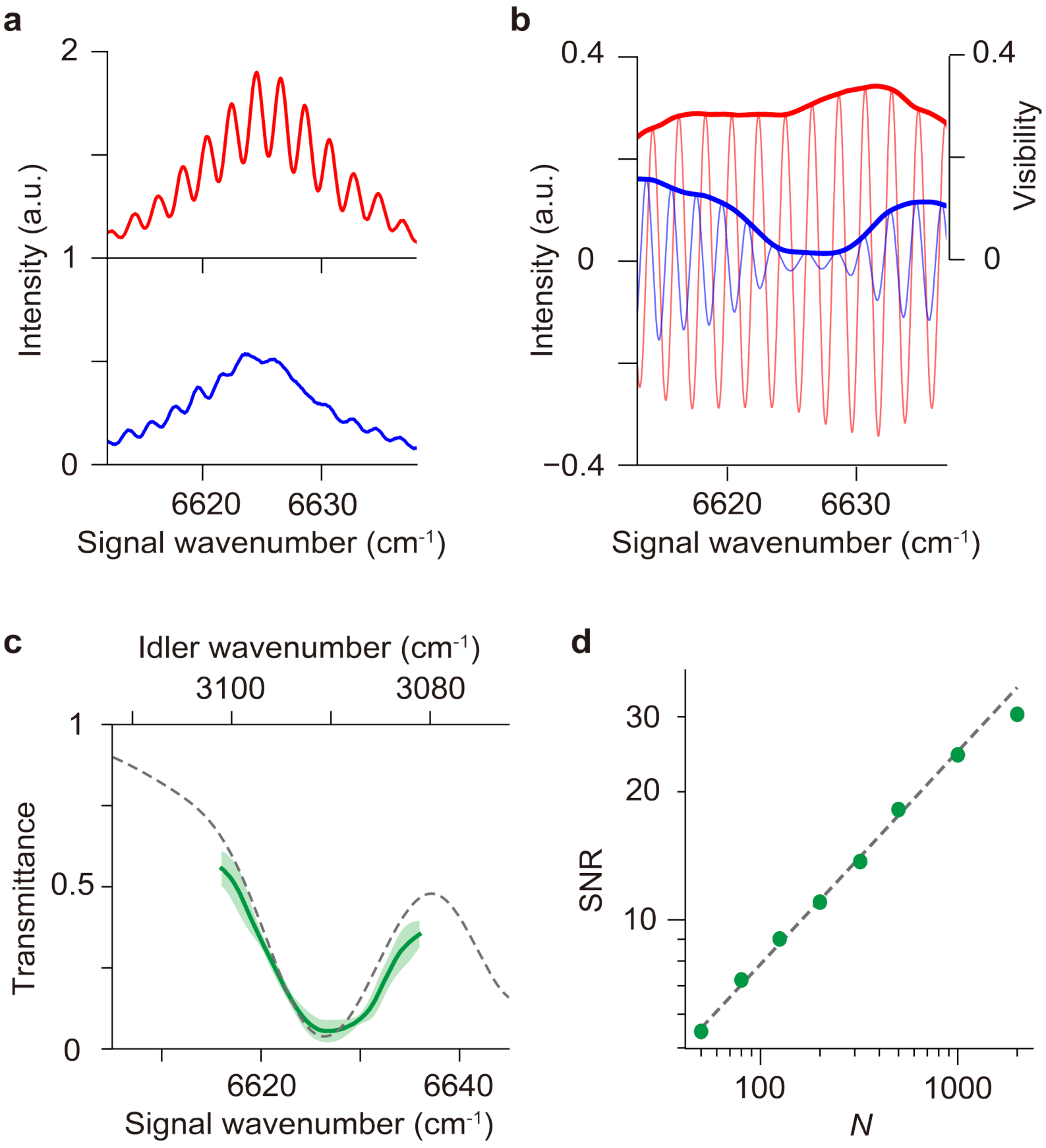

**Fig. 4: High-speed MIR transmittance spectroscopy.** (a) Time-stretch signal spectra measured without (red) and with (blue) liquid benzene, each averaged over 1,000 consecutive pulses at a repetition rate of 1 MHz. (b) Corresponding normalized interference waveforms (thin curves) and fringe visibilities extracted using the Hilbert transform (thick curves). Red and blue indicate measurements without and with the sample, respectively. (c) Retrieved MIR transmittance spectrum (green solid curve) with ±1 standard deviation (shaded area), calculated from eight independently retrieved spectra, each averaged over 1,000 pulses. The dashed gray curve represents the reference FTIR spectrum convolved to an estimated spectral resolution of 6.2 cm$^{-1}$. (d) SNR as a function of the number of averaged pulses, $N$. The green points represent experimental values, and the dashed line shows a square-root fit of $0.79\sqrt{N}$.

**Broadband MIR transmittance spectroscopy**

Finally, we demonstrated the broadband measurement capability of TS-MIR-SUP using the 10-mm-long PPLN crystal, which provides a wider phase-matching bandwidth than the 50-mm-long crystal used for high-speed measurements. Figures 5a and b show the retrieved transmittance spectra of a MIR bandpass filter and an acrylic plate, respectively, measured with a poling period of 30.49 µm. The spectra were obtained by averaging 2,500 pulses at a repetition rate of 50 kHz. The crystal temperature was set to 40 °C for the bandpass filter and to 80 °C and 30 °C for the acrylic plate. The retrieved spectra agree well with the corresponding FTIR reference spectra over a spectral range of approximately 130 cm$^{-1}$ for each measurement condition, with SNRs of 20 and 24 for the bandpass filter and acrylic plate, respectively.

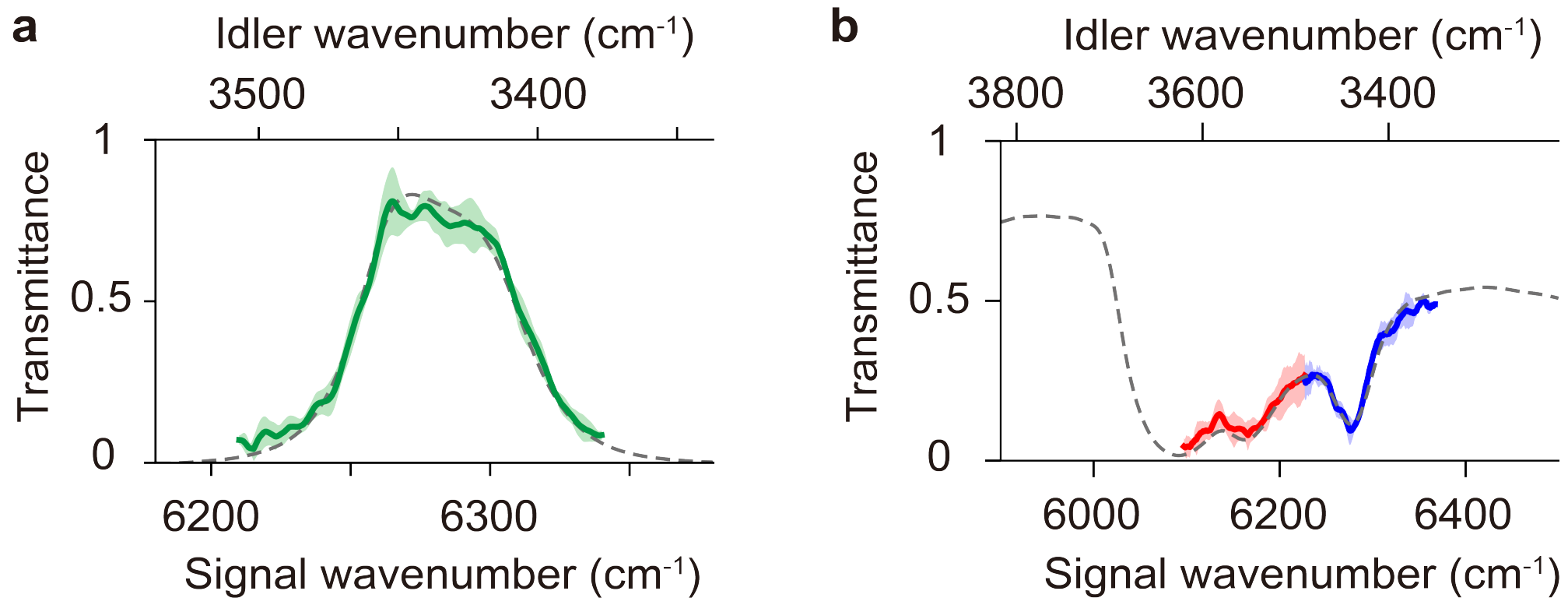

**Fig. 5: Broadband MIR transmittance spectroscopy.** Retrieved transmittance spectra of (a) a MIR bandpass filter (green) and (b) an acrylic plate (red and blue), measured using a 10-mm-long PPLN crystal at a repetition rate of 50 kHz. The spectra were obtained by averaging 2,500 pulses for both samples. The crystal temperature was 40 °C for (a) and 80 °C (red) and 30 °C (blue) for (b). Gray dashed curves represent the corresponding FTIR reference spectra.

## Discussion

### Spectral acquisition rate

The spectral acquisition rate of TS-MIR-SUP is determined by the number of pulses required to achieve a target SNR in the retrieved MIR transmittance spectrum. Although individual NIR signal spectra can be acquired at the pump repetition rate, averaging is necessary to suppress intensity fluctuations arising from the thermal photon-number statistics of PDC light. Increasing the pump repetition rate could therefore improve the spectral acquisition rate, provided sufficient signal intensity is maintained. However, at higher repetition rates, the available pump pulse energy decreases under a limited average pump power, reducing the parametric gain and detected signal intensity. In our system, the repetition rates were set to 1 MHz and 50 kHz for the 50- and 10-mm-long PPLN crystals, respectively, with corresponding pump pulse energies of 127 and 526 nJ.

Improving the efficiency of PDC offers a route to higher repetition rates without sacrificing signal intensity. One promising approach is to use a nonlinear waveguide [25], which could enhance parametric conversion efficiency, thereby reducing the required pump pulse energy. Another possibility is to introduce a seed beam to increase the detectable signal intensity. For example, Dong et al. demonstrated MIR-SUP with idler seeding using a wavelength-tunable MIR quantum cascade laser (QCL) [26]. Although idler seeding could relax the pump-energy requirements, it introduces an additional MIR light source. Waveguide-based enhancement may therefore be particularly attractive for increasing the acquisition rate while retaining the simple NIR semiconductor-laser-based architecture demonstrated here.

**Spectral coverage**

The accessible MIR wavelength range of TS-MIR-SUP can be tuned from 2.7 to 3.4 μm by adjusting the phase-matching conditions of the PPLN crystal (Supplementary Note 4). The instantaneous spectral bandwidth could be further extended by employing chirped quasi-phase-matching gratings in nonlinear crystals or waveguides, potentially enabling broadband measurements spanning several hundred or even a few thousand $cm^{-1}$ [14,16]. Another approach is to shift the pump wavelength from 1029 to 822 nm, which could enable broadband signal and idler generation around 1.1 and 3.3 μm, respectively, by exploiting group-velocity matching in MgO-doped PPLN [27]. The availability of semiconductor lasers at a wide range of wavelengths provides flexibility in selecting suitable pump wavelengths while retaining a simple laser-source architecture. Such an extension would require appropriate optical amplification to achieve sufficient parametric gain.

**Spectral resolution**

The spectral resolution of TS-MIR-SUP in the liquid benzene measurement was estimated from the RF bandwidth available for retrieving sample spectral information from the interference fringes. In the present visibility-retrieval procedure, the non-interference background was estimated using a low-pass filter with a cutoff frequency of 0.85 GHz. Meanwhile, the finite temporal correlation response attenuates higher RF frequencies, with an estimated upper usable frequency of approximately 1.65 GHz. With a fringe carrier frequency of 1.25 GHz, these constraints yield a usable modulation bandwidth of approximately 0.40 GHz on either side of the carrier. Using the measured chirp rate of 2.5 $cm^{-1}$ $ns^{-1}$ at a wavelength of 1.51 μm, this corresponds to a nominal spectral resolution of approximately 6.2 $cm^{-1}$. This estimate depends on the present visibility-retrieval procedure and should not be regarded as a fundamental resolution limit of TS-MIR-SUP. Further improvement could be achieved by reducing the temporal correlation width through narrowing of the pump spectral linewidth, for example by suppressing self-phase modulation in the YDFA using a large-mode-area fiber. With sufficiently high parametric gain, longer pump pulses with narrower spectral linewidths could potentially enable gas-phase absorption spectroscopy.

**Performance comparison**

We first compare TS-MIR-SUP with a previously reported low-gain MIR-SUP system using a grating spectrometer [15], which, to the best of our knowledge, achieved the highest spectral acquisition rate among prior MIR-SUP demonstrations. The previous system achieved a measurement time of 10 ms over a spectral bandwidth of 270 $cm^{-1}$ with a resolution of 6 $cm^{-1}$ and an SNR of 34. In comparison, our system achieves a measurement time of 1 ms over a bandwidth of 20 $cm^{-1}$ with an estimated resolution of 6.2 $cm^{-1}$ and an SNR of 24. Although the measurement conditions and SNR definitions differ, our approach demonstrates a 10-fold reduction in acquisition time. In the previous system, the acquisition rate was limited by the low photon flux of SPDC and the readout speed of the grating spectrometer. These limitations could be mitigated by high-gain operation or seeding and the use of a faster spectrometer, respectively. In contrast, time-stretch detection enables pulse-resolved spectral acquisition at the pump repetition rate, offering a route to further improvements in measurement speed as higher repetition rates become accessible.

Another approach to high-speed MIR spectroscopy is up-conversion spectroscopy [28–31], in which MIR light is frequency-converted to the VIS/NIR region for detection. Unlike TS-MIR-SUP, which requires averaging to suppress the intensity fluctuations of PDC light, up-conversion spectroscopy can provide single-shot spectral measurements with relatively small pulse-to-pulse intensity fluctuations [31]. However, broadband implementations typically require a femtosecond MIR source and an additional optical field for frequency conversion, increasing system complexity. In contrast, TS-MIR-SUP generates broadband, frequency-correlated MIR and NIR light directly from a narrowband pump, allowing implementation with a fiber-amplified semiconductor laser.

MIR ghost spectroscopy (MIR-GS) using bright twin beams, which we recently developed, provides another approach to high-speed MIR spectroscopy [32]. This technique exploits the correlated pulse-to-pulse spectral fluctuations of high-gain PDC light by combining time-stretch detection of the NIR signal spectrum with spectrally integrated detection of the MIR idler. Because MIR-GS requires only a single PDC process, it avoids the need for interferometric phase stability. However, it requires an MIR bucket detector, whose noise can limit measurement sensitivity. In contrast, TS-MIR-SUP enables the retrieval of both MIR absorption and phase spectra using only NIR detection. Furthermore, PSA in the second nonlinear interaction amplifies the signal generated in the first PDC process, potentially allowing operation with lower initial photon flux. This could facilitate higher repetition rates, broader spectral coverage using shorter crystals, and the use of nonlinear materials with smaller nonlinear coefficients.

**Future prospects**

The high spectral acquisition rate of TS-MIR-SUP offers opportunities for monitoring rapid and nonrepetitive processes. For example, millisecond-scale, chemically selective measurements could enable real-time analysis of microfluidic flows, with potential applications in particle identification and sorting. The approach could also be extended to MIR hyperspectral imaging [33,34], where high-speed spectral acquisition could substantially reduce the time required for spatial scanning. Another potential application is MIR optical coherence tomography [35–37], where time-stretch detection could accelerate spectral readout and volumetric imaging. Furthermore, the pulsed operation of TS-MIR-SUP could enable time-resolved MIR spectroscopy, including pump–probe measurements [38].

Extending the accessible wavelength range into the molecular fingerprint region could broaden the applications of TS-MIR-SUP, including time-resolved studies of biological processes such as protein folding [39]. MIR-SUP has already been demonstrated in this region using $AgGaS_2$ (AGS) [40,41], $BaGa_2GeSe_6$ (BGGSE) [42], and GaSe [43] crystals in the low-gain regime. Adapting these materials to high-gain operation would require sufficient parametric amplification. Alternatively, orientation-patterned GaP (OP-GaP) [44], which offers a large nonlinear coefficient and supports type-0 quasi-phase matching without birefringent spatial walk-off, could be a promising candidate for extending high-speed MIR-SUP into this spectral region.

Finally, the pulse-resolved spectral detection enabled by time-stretch spectroscopy provides access to the intensity fluctuations inherent in high-gain PDC light. Although these fluctuations currently necessitate averaging over multiple pulses, their known statistical properties could potentially be exploited to improve spectral reconstruction.

For example, statistical estimation methods incorporating the thermal photon-number distribution into the measurement model may enable more efficient retrieval of sample transmittance than simple averaging, potentially reducing the number of pulses required to achieve a given estimation precision.

## Methods

### TS-MIR-SUP system

The pump source consisted of a distributed-feedback laser diode (QD Laser) operated in the gain-switching regime by direct electrical modulation using a function generator. The laser generated approximately 50-ps pulses centered at 1029 nm, which were subsequently amplified to sub-μJ energies using a YDFA module (Keopsys). The amplified pump pulses had spectral linewidths (FWHM) of 0.14 nm (1.3 $cm^{-1}$) and 0.28 nm (2.6 $cm^{-1}$) for the 1-MHz and 50-kHz measurements, respectively. For the 50-kHz measurements, the pump beam was collimated into free space, and its power was adjusted using a half-wave plate (HWP) and a polarizing beam splitter (PBS). A small fraction of the pump light was extracted from the other PBS port to generate a trigger signal for the time-stretch waveforms. For the 1-MHz measurements, the HWP and PBS were removed to minimize pump losses. Instead, residual pump light coupled into the fiber together with the signal pulses after the interferometer was separated using a wavelength-division multiplexer (WDM) to provide the trigger signal. In both configurations, the extracted pump pulses were detected using a 10-GHz InGaAs photodetector.

After polarization adjustment using an HWP and a quarter-wave plate (QWP), the pump pulses were focused into a 5 mol% MgO-doped PPLN crystal under type-0 phase-matching conditions. Two crystals were used: a 50-mm-long fan-out PPLN crystal (HC Photonics) with poling periods ranging from 27.5 to 31.6 μm for the 1-MHz measurements, and a 10-mm-long PPLN crystal (HC Photonics) with discrete poling periods of 29.52, 29.98, and 30.49 μm for the 50-kHz measurements. The pump pulses were focused using lenses with focal lengths of 300 and 100 mm for the 50- and 10-mm-long crystals, respectively. Both crystals were mounted in temperature-controlled ovens.

The SU(1,1) interferometer was implemented in a folded configuration. After the first passage through the PPLN crystal, the pump, signal, and idler beams were separated using a dichroic mirror. The pump and signal beams were reflected back into the crystal by a concave mirror, while the idler beam was collimated using a $CaF_2$ lens and reflected by an end mirror. For the 50- and 10-mm-long crystal configurations, both the concave mirror and $CaF_2$ lens had focal lengths of 150 and 100 mm, respectively. The end mirror was mounted on a motorized translation stage to adjust the optical path difference (OPD) between the pump–signal and idler arms.

The samples consisted of liquid benzene with a thickness of 25 μm, a 1-mm-thick MIR bandpass filter (Multi IR Technology), and a 0.2-mm-thick acrylic plate (CLAREX), each placed in the idler arm. Sample insertion altered the OPD and idler-beam alignment, causing additional reductions in fringe visibility. To compensate for these changes, the end-mirror position was adjusted to restore the OPD, and its tilt was optimized to maximize the fringe visibility. Reference spectra were measured using a commercial FTIR spectrometer (JASCO).

After the second passage through the PPLN crystal, the signal beam was separated from the pump and idler beams using a dichroic mirror, and residual pump light was suppressed using a long-pass filter. The signal beam was then coupled into a 1-m-long telecom-band single-mode fiber, followed by a 10-km dispersion-compensating fiber (YOFC). The time-stretch waveforms were detected using a 10-GHz InGaAs photodetector (Thorlabs) and digitized using a 16-GHz oscilloscope (Teledyne LeCroy) at a sampling rate of 80 GSa/s.

**Time-stretch spectral calibration**

The time-to-wavenumber mapping of the time-stretch waveforms was calibrated following a previously demonstrated procedure [45]. First, the relative optical delay between the pump–signal and idler arms was determined to be 16.1 ps from the spectral fringe spacing measured using an optical spectrum analyzer (OSA) (ANDO). The temporal phase of the time-stretch interference waveform was then extracted using the Hilbert transform, and its time derivative was used to calculate the instantaneous RF fringe frequency. Combined with the measured optical delay, this frequency provided the chirp rate as a function of time. Integration of the chirp rate yielded the relative time-to-wavenumber mapping. Finally, the absolute wavenumber offset was determined by matching the interference fringe peaks in the time-stretch and OSA spectra, with the sample absorption features used to resolve ambiguities in fringe peak correspondence. The group-velocity dispersion (GVD) and third-order dispersion (TOD) coefficients of the DCF were also determined from the time dependence of the chirp rate, yielding 210.4 $ps^2$ $km^{-1}$ (corresponding to a chirp rate of 2.5 $cm^{-1}$ $ns^{-1}$) and −2.8 $ps^3$ $km^{-1}$, respectively, at a signal wavelength of 1.51 μm.

**Visibility extraction and transmittance retrieval**

The sample transmittance was retrieved from the fringe visibilities of the time-stretch spectra measured with and without the sample. For each averaged spectrum, the non-interference background was first estimated using a zero-phase low-pass filter. The measured waveform was then divided by the background, and unity was subtracted to obtain the normalized interference waveform. A second low-pass filter was applied to suppress high-frequency noise while preserving the interference fringes. The fringe visibility at each signal wavenumber was extracted from the envelope of the normalized waveform using the Hilbert transform. In the high-gain regime, the fringe visibility $V$ is given by $V = \frac{2T}{1+T^2}$, where $T$ denotes the ratio of the idler-to-signal amplitude attenuation along their respective optical paths between the first and second PDC processes [20]. The visibilities measured with and without the sample were individually converted to the corresponding values of $T$. Their ratio eliminates the optical losses common to both measurements, yielding the round-trip amplitude transmittance introduced by the sample. Because the idler beam passes through the sample twice in the folded interferometer, this ratio corresponds to the single-pass intensity transmittance of the sample.

**Funding**

This work was supported by JSPS KAKENHI (23H00273, 25H01386, T.I., 25K23541, K.H.), JST FOREST Program (JPMJFR236C, T.I.), RIKEN TRIP initiative (HIKARI-COOL Tokyo, T.I.), and JST PRESTO (JPMJPR2457, K.H.)

**Acknowledgements**

The authors acknowledge Ryoichi Horisaki and Akira Kawai for inspiring discussions, Kazunori Naganuma for providing the gain-switched laser diode and YDFA, and Zicong Xu and Zhihao Deng for providing constructive comments on the manuscript.

**Author contributions**

K.H. and T.I. conceived the project. M.S. and K.H. designed and built the optical setup. M.S. and T.Y. conducted the initial experiments to verify PDC generation. M.S. acquired all experimental data, developed the analysis code, and performed the data analysis. K.H. provided technical advice and suggestions throughout the project. M.S., K.H., and T.I. discussed the results. K.H. and T.I. supervised the work. M.S., K.H., and T.I. wrote the manuscript.

**Competing Interests**

M.S., K.H. and T.I. are named inventors on a patent application related to the TS-MIR-SUP system.

**Data availability**

The data presented in the manuscript are available from the corresponding author upon reasonable request.